\documentclass[11pt]{article}
\usepackage{moriond,epsfig}

\def\be{\begin{equation}}
\def\ee{\end{equation}}
\def\bea{\begin{eqnarray}}
\def\eea{\end{eqnarray}}

\begin{document}
\title{STOCHASTIC FIELD THEORY FOR TRANSPORT STATISTICS
IN DIFFUSIVE SYSTEMS}

\author{EUGENE V. SUKHORUKOV,
ANDREW N. JORDAN, SEBASTIAN PILGRAM}

\address{D\'epartement de Physique Th\'eorique, Universit\'e de Gen\`eve,
CH-1211 Gen\`eve 4, Switzerland}

\maketitle
\abstracts{
We present a field theory for the statistics of charge and current 
fluctuations in diffusive systems. The cumulant generating function
is given by the saddle-point solution for the action of this field theory.
The action depends on two parameters only: the local diffusion and 
noise coefficients, which naturally leads to the universality of 
the transport statistics for a wide class of multi-dimensional 
diffusive models. Our theory can be applied to semi-classical
mesoscopic systems, as well as beyond mesoscopic physics.
}

\section{Introduction}
In this short paper, we present the essential part of our theory\,\cite{us1}
for the statistics of current and density fluctuations in non-equilibrium 
diffusive systems. The theory is based on the stochastic path integral  
approach to the statistics of fluctuations in networks,\cite{us2} and represents 
an alternative to quantum methods\,\cite{LevitovStatistics,Nazarov1}
for the evaluation of the full counting statistics (FCS) of the transmitted charge, 
as well as to classical cascade correction method\,\cite{Nagaev1} 
for the evaluation of current cumulants.

The building blocks of our theory are a separation of the time scales 
(sources of noise are Markovian) and a saddle-point approximation of the resulting
functional integral. 
The separation of variables into fast currents and slow varying generalized
charges suggests the network description of a system introduced in  
Sec.\ \ref{SPI}. In Sec.\ \ref{CL} we consider a large number 
of nodes and derive the continuum limit for the stochastic path integral, 
which turns out
to be a stochastic field theory with an action being a function 
of only two quantities, the diffusion $D$ and noise $F$ functionals
of the charge density $\rho$. This naturally 
leads to the universality of the FCS not only for the mesoscopic diffusive
conductor [see Eq.\ (\ref{ans})], but for a whole class of multi-dimensional 
diffusive models (see also Ref.\ [1] for details).
 
We would like to stress that due to its essentially classical construction, our 
stochastic field theory is not limited to mesoscopic physics,\cite{pilgram} 
but can also be applied to reaction-diffusion systems,\cite{Kamenev} 
symmetric exclusion processes,\cite{derrida} and many other areas 
of classical stochastic processes.\cite{gardiner} 

\section{Stochastic Path Integral for a Network}
\label{SPI}

Consider a network with the state of each node $\alpha$ described by one charge 
$Q_\alpha$  (${\bf Q}$ is the charge vector describing the charge
state of the network).  The node's state may be changed by transport: 
flow of charges between nodes takes place via the connectors 
carrying currents $I_{\alpha\beta}$ from node $\alpha$ to node $\beta$.  
The rate of change of
these charges $Q_{\alpha}$ is given by 
\be \dot Q_{\alpha}=\sum_{\beta}I_{\alpha\beta}\, ,\qquad
P_{\alpha\beta}(I_{\alpha\beta})=\int\frac{d\lambda_{\alpha\beta}}{2\pi}
\exp\{-itI_{\alpha\beta}\lambda_{\alpha\beta}+tH_{\alpha\beta}
(\lambda_{\alpha\beta})\}\,,
\label{cons}
\ee 
where $P_{\alpha\beta}$ is the probability distribution
of the fast current between nodes $\alpha$ and $\beta$, and
$H_{\alpha\beta}$ is the current cumulant generating function.
The fact that $P_{\alpha\beta}$ also 
depends on the charges ${\bf Q}$ is one source of the
difficulty of the problem.

Assuming that the generators $H_{\alpha\beta}$ of the fluctuating
currents $I_{\alpha\beta}$ are known, we seek an evolution of the probability
distribution $\Gamma({\bf Q},t)$ of the set of charges ${\bf Q}$ for a given
initial condition $\Gamma({\bf Q},0)$. In other words, one has to find the
conditional probability (which we refer to as the evolution operator) $U({\bf
Q},{\bf Q}',t)$ such that 
\be \Gamma({\bf Q},t)=\int d{\bf Q}'\, U({\bf
Q},{\bf Q}',t)\Gamma({\bf Q}',0)\, .
\label{gamma}
\ee 
We assume that there is a separation of time scales, $\tau_0 \ll \tau_C$,
between the correlation time of current fluctuations, $\tau_0$, and the slow
relaxation time of charges in the nodes, $\tau_C$.  
In Ref.\ [2] we have used the separation of time scales to derive 
a stochastic path integral representation for the evolution operator,
\bea 
U({\bf Q}_f,{\bf Q}_i,t) 
&=&\int\! {\cal D}{\bf Q}{\cal D}{\bf \Lambda}
\exp\{S({\bf Q},{\bf \Lambda})\}, 
\label{path1}\\  
S({\bf Q},{\bf \Lambda}) 
&=&\int_{0}^{t} dt'[-i 
{\bf \Lambda}\cdot {\dot {\bf Q}}
+(1/2)\sum_{\alpha\beta}H_{\alpha\beta} 
({\bf Q}, \lambda_\alpha-\lambda_\beta)]. 
\label{action1} 
\eea
The variables $\lambda_\alpha$ are auxiliary variables 
for every node that impose charge conservation in the network.

\section{Continuum Limit}
\label{CL}

\begin{figure}[t]
\epsfxsize=3in
\center{\epsfbox{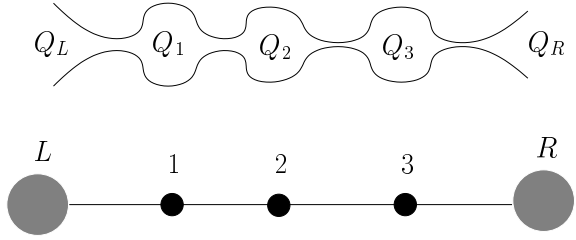}}
\caption{A one dimensional lattice of nodes connected on both ends
to reservoirs.  This situation could represent a series of
mesoscopic chaotic cavities connected by point contacts.}
\label{lattice}
\end{figure}

From the stochastic network, it is straightforward to go
to spatially continuous systems as the spacing between the nodes is taken to
zero. Consider a series of identical, equidistant nodes separated by a distance
$\Delta z$.  This nodal chain could represent a chain of chaotic cavities,
Fig.\ \ref{lattice}, in a mesoscopic context.\cite{Oberholzer1,Ob2} The sum
over $\alpha$ and $\beta$ becomes a sum over each node in space connected to
its neighbors.  The action for this arrangement is
\be
S = \int_0^t dt' \sum_{\alpha} \{-\lambda_\alpha {\dot Q}_\alpha
+H(Q_{\alpha}, Q_{\alpha-1};\lambda_{\alpha}-\lambda_{\alpha-1})\}\; ,
\label{act1}
\ee
where for simplicity we have chosen real counting variables,
$i\lambda_\alpha\rightarrow \lambda_\alpha$.  The only constraint made on $H$
is that probability is conserved, $H(\lambda_{\alpha}-\lambda_{\alpha-1})=0$
for $\lambda_{\alpha}=\lambda_{\alpha-1}$.  We now derive a lattice field
theory by formally expanding $H$ in $\lambda_{\alpha}-\lambda_{\alpha-1}$ and
$Q_{\alpha}-Q_{\alpha-1}$.  Only differences of the counting variables will
appear in the series expansion, while we must keep the full $Q$ dependence of
the Hamiltonian.  If there are $N \gg 1$ nodes in the lattice, for fixed
boundary conditions the difference between adjacent variables,
$\lambda_{\alpha}-\lambda_{\alpha-1}$ and $Q_{\alpha}-Q_{\alpha-1}$ will be of
order $1/N$, and therefore provides a good expansion parameter.  The expansion
of the Hamiltonian (\ref{act1}) to second order in the difference variables
gives
\be
H =  \frac{\partial H}{\partial
 \lambda_{\alpha}}\, (\lambda_{\alpha}-\lambda_{\alpha-1}) + \frac{1}{2}
   \, \frac{\partial^2 H}{\partial
 \lambda_{\alpha}^2} (\lambda_{\alpha}-\lambda_{\alpha-1})^2 + 
\frac{\partial^2 H}{\partial
    Q_{\alpha}\partial \lambda_{\alpha}} (Q_{\alpha}-Q_{\alpha-1})
  (\lambda_{\alpha}-\lambda_{\alpha-1})\, , 
\label{expandh}
\ee
where the expansion coefficients are evaluated at
$\lambda_{\alpha}=\lambda_{\alpha-1}$ and $Q_{\alpha}=Q_{\alpha-1}$ and are
functions of $Q_{\alpha-1}$.  Terms involving only differences of
$Q_\alpha-Q_{\alpha-1}$ are zero because
$H(\lambda_{\alpha}-\lambda_{\alpha-1})=0$ for
$\lambda_{\alpha}=\lambda_{\alpha-1}$.  All terms in Eq.~(\ref{expandh}) need
explanation.  First, the expression ${\partial H}/{\partial\lambda_\alpha}$ is
the local current at zero bias (because the charges in adjacent nodes are
equal) which will usually be zero.  There may be exceptional circumstances
where this term should be kept, but we do not consider them here.
The term $ \partial^2 H/{\partial Q_{\alpha} \partial\lambda_{\alpha}} =
-G(Q_{\alpha-1})$ is the linear response of the current to a charge
difference.  Hence, $G$ is the generalized conductance\,\cite{cap} of the
connector between nodes $\alpha$ and $\alpha-1$.  $\partial^2 H/{\partial
\lambda_{\alpha}^2} = C(Q_{\alpha-1})$ is the current noise through the same
connector because $H$ is the generator of current cumulants.

We are now in a position to take the continuum limit by replacing the node
index $\alpha$ with a coordinate $z$, introducing the fields $Q(z),
\lambda(z)$, and making the expansions
\begin{eqnarray}
\lambda_{\alpha}-\lambda_{\alpha-1}\!\to\! \lambda'\Delta z+
(1/2)\lambda''(\Delta z)^2+{\cal O}(\Delta z)^3,\, \label{expa}\\
Q_{\alpha}- Q_{\alpha-1}
\!\to\! Q'\Delta z+
(1/2) Q''(\Delta z)^2+{\cal O}(\Delta z)^3.\,
\label{expb}
\end{eqnarray}
The action may now be written in terms of intensive fields
by scaling away $\Delta z$,
\be
H\rightarrow   h (\rho,\lambda) \Delta z,\quad 
Q_\alpha \rightarrow \rho(z) \Delta z,\quad 
G_\alpha (\Delta z)^2 \rightarrow D(\rho),\quad 
C_\alpha \Delta z \rightarrow F(\rho)\, ,
\label{redef}
\ee
and taking the limit $\sum_\alpha H \rightarrow \int dz h (\rho, \lambda)$.
One may check that expanding the Hamiltonian to higher than second order in
$\Delta z$ will result in terms suppressed by powers of $\Delta z/L$ and
consequently vanish as $\Delta z\rightarrow 0$.  

These considerations leave the one dimensional action as
\be
S = -\int_0^t dt'\int_0^L dz \left [\lambda  {\dot \rho}+  D\, \rho' \lambda'
- \frac{1}{2}F\, (\lambda')^2 
\right ]\, .
\label{act1d}
\ee
Here $D$ is the local diffusion constant and $F$ is the local noise density
which are discussed in detail below.  It is very important that these two
functionals $D,F$ are all that is needed to calculate current statistics.
Classical field equations may be obtained by taking functional derivatives of
the action with respect to the charge and counting fields:
\be
{\dot \lambda} =-\frac{1}{2}\frac{\delta F}{\delta \rho} \,(\lambda')^2
- D\lambda'' , \qquad
  {\dot \rho} = [-F  \lambda' + D \rho']'\, .
\label{eom1}
\ee
We have to solve these coupled differential
equations subject to the boundary conditions
%
%
$\rho(t, 0) =  \rho_L(t)$, $\rho(t, L) =  \rho_R(t)$,
$\lambda(t, 0) =  \lambda_L(t)$, and $\lambda(t, L) =  \lambda_R(t)$.
Functions $\lambda_L(t)$ and $\lambda_R(t)$ are the
counting variables of the absorbed charges at the left and right end
of the system. Once Eqs.~(\ref{eom1}) are solved, 
the solutions $\rho(z,t)$ and $\lambda(z,t)$ should be
substituted back into the action (\ref{act1d}) and integrated over time and
space.  The resulting function,
$S_{sp}[\rho_L(t),\rho_R(t),\lambda_L(t),\lambda_R(t),t,L]$ is the generating
function for time-dependent cumulants of the current distribution.  Often, the
relevant experimental quantities are the stationary cumulants.  These are
given by neglecting the time dependence, finding static solutions, ${\dot
\rho} ={\dot \lambda}=0$, and imposing static boundary conditions.  We can 
also introduce sources $\int dtdz\,\chi(z,t)\rho(z,t)$ 
and calculate density correlation functions.

To justify the saddle-point approximation, it is useful
to define dimensionless variables.  The boundary conditions $\rho_L$, and
$\rho_R$ provide the charge density scale $\rho_0$ in the problem, so we
define $\rho(z) =\rho_0 f(z)$, where $f\sim 1$ is an occupation.  We
furthermore rescale $z\to Lz$, and $t\to\tau_D t$, where $\tau_D=L^2/D$ is the
diffusion time, thus obtaining
\be
S= -L \rho_0\int_0^{\,t}\! dt'\!\int_0^1 dz' 
\left [\lambda  {\dot f}+  f' \lambda'-\frac{F}{2D\rho_0}(\lambda')^2 
\right ]\, .
\label{act1d2}
\ee
We assume that the combination ${F}/{D\rho_0}$ is of order 1.  From
Eq.~(\ref{act1d2}), the dimensionless large parameter is $\gamma=\rho_0L\gg
1$, i.e.\ the number of transporting charge carriers.  The saddle-point 
contribution is of order $\gamma t/\tau_D$, while the
fluctuation contribution is of order $t/\tau_D$.

Repeating these steps in multiple dimensions 
yields the action
\be
S = -\int_0^{\,t} \! dt'\! \int_\Omega d{\bf r} \; [\,  {\lambda}
{\dot \rho} + \nabla{\lambda} \, {\hat D}\, \nabla{\rho} -
(1/2)\nabla{\lambda}\, {\hat F}\, \nabla{\lambda}
\, ]\; ,
\label{fieldaction}
\ee
where ${\hat F}$ and ${\hat D}$ are general matrix
functions of the density ${\rho}$ and coordinate ${\bf r}$ which should
be interpreted as noise and diffusion matrices. 

As in any field theory, symmetries of the action play an important
role because they lead to conserved quantities.  We first note that
the Hamiltonian $h(\rho, \nabla \rho, \nabla { \lambda})$ is a
functional of $\nabla {\lambda}$ alone with no ${\lambda}$ dependence.
This symmetry is analogous to gauge invariance, and leads to the
equation of motion 
\be
{\dot \rho} + \nabla \cdot {\bf j}=0\, ,\quad
{\bf j}=-{\hat D}\nabla \rho + {\hat F}\nabla {\lambda}\, ,
\label{conslaw1}
\ee  
which can
be interpreted as conservation of the conditional current density ${\bf j}$. 
The next symmetry is
related to the invariance under a shift in the space and time
coordinates $\{{\bf \delta \bf r}, \delta t\}$. This symmetry leads to
equations analogous to the conservation of the local energy/momentum
tensor.\cite{landau} For the
stationary limit (where $\dot \rho$ and $\dot \lambda$ vanish) and for
symmetric diffusion and noise tensors, the conservation law is 
relativly simple and is given by
\be
\sum_m \nabla_m T_{mn} = 0\, ,\quad 
T_{mn} = j_m (\nabla_n \lambda) - (\nabla_n \rho)\, (D \nabla \lambda)_m -
h\, \delta_{mn} \, .
\label{conslaw2}
\ee
For the special case of a one dimensional geometry, 
the Hamiltonian itself is the conserved quantity (see Sec.~\ref{FCS}).
 
\section{FCS of Diffusive Systems}
\label{FCS}

We first consider the general 1D field theory with the action
(\ref{act1d}), and then demonstrating our solution for the FCS 
of the mesoscopic diffusive wire specifically. 
In the stationary limit, ${\dot \rho} = {\dot \lambda}=0$, the
action can be written as
\be
S=t\int\limits_0^Ldz \left[-D\rho'\lambda'+\frac{1}{2}F(\lambda')^2\right].
\label{action-stat1}
\ee
The stationary saddle-point equations
\begin{equation}
(F\lambda'-D\rho')'=0, \quad
2D\lambda''+\frac{\delta F}{\delta\rho}\,(\lambda')^2=0,
\label{spe12}
\end{equation}
can be partially integrated leading to the 
following two equations:
\be
D\rho'=\pm\sqrt{{\cal I}^2-2{\cal H}F},
\label{sol1}
\ee
\be
\lambda'=2{\cal H}/({\cal I}-D\rho').
\label{sol2}
\ee
The two integration constants ${\cal I}=-D\rho'+F\lambda'$ and ${\cal
H}=-D\rho'\lambda'+(F/2)(\lambda')^2$ are the 
conserved (conditional) current and
the Hamiltonian density, respectively.  These conservation laws follow from
the symmetries of our $1D$ field theory
[see Eqs.\ (\ref{conslaw1}) and (\ref{conslaw2}) and the surrounding discussion].  Thus we
obtain the following result for the action (\ref{action-stat1}),
\be
S=tL{\cal H}.
\label{action-stat2}
\ee

The Eqs.\ (\ref{sol1}-\ref{action-stat2}) represent
the formal solution of the FCS problem for 1D diffusion models with $D(\rho)$
and $F(\rho)$ being arbitrary functions of $\rho$.  The following procedure
has to be done in order to obtain the cumulant generating function $S(\chi)$
of the transmitted charge:
(i) The differential equation (\ref{sol1}) has to be solved for $\rho(z)$ with the
boundary conditions $\rho(z)|_{z=0}=\rho_L$ and $\rho(z)|_{z=L}=\rho_R$. The
constant ${\cal I}$ should be expressed through the constants $\rho_L$,
$\rho_R$, and ${\cal H}$. (ii)
Next, $\rho(z)$ is substituted into Eq.~(\ref{sol2}) which is integrated to
obtain $\lambda(z)$ with the boundary conditions $\lambda_L=0$ and
$\lambda_R=\chi$.
(iii) Finally, using the solution for $\lambda(z)$ the constant ${\cal H}$ is
expressed in terms of $\rho_L$, $\rho_R$, $\chi$, and substituted into the
action (\ref{action-stat2}).
We note that by expressing ${\cal H}$ and $\chi$ in terms of ${\cal I}$, 
we may also formally obtain the logarithm of the current distribution, 
\be
\ln P(I)=S({\cal I})-t{\cal I}\chi({\cal I}),\quad {\cal I}\to I,
\label{LNPI}
\ee
as a result of the stationary phase approximation for the integral $P(I)=\int
d\chi\exp[S(\chi)-tI\chi]$ and because $\partial {\cal H}/\partial\chi={\cal
I}/L$.

As an example of the 1D field theory, we consider the FCS of the electron
charge transmitted through the mesoscopic diffusive wire. When the potential
difference $\Delta\mu=\mu_L-\mu_R>0$ is applied to the wire, the electrons
flow from the left lead to the right lead with the average current
$I_0=e^{-1}G\Delta\mu$, where $G$ is the conductance of the wire. The elastic
electron scattering causes non-equilibrium fluctuations of the current.  At
zero temperature, and for noninteracting electrons (the cold electron regime),
the FCS of the transmitted charge has been studied in Refs.\ [14] and 
[4] using quantum-mechanical methods with the following result 
for the generating function of cumulants of the dimensionless charge $Q/e$:
\be
S(\chi) = (tI_0/e)\,{\rm arcsinh}^2 \left[\sqrt{\exp(\chi)-1}\,\right].
\label{ans}
\ee
Here we will rederive this result using our classical method.
\begin{figure}[t]
 \vspace{20pt}
  \centerline{\hbox{ \hspace{0.0in} 
    \epsfxsize=3.1in
    \epsffile{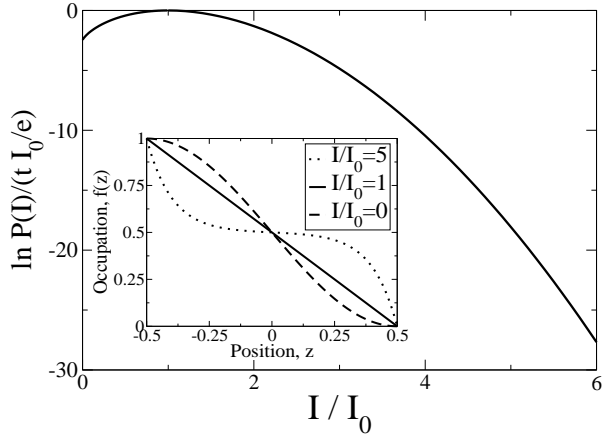}
}}
  \caption{The logarithm of the distribution of the current
  through a mesoscopic diffusive wire as a function of the ratio 
  $I/I_0$ of the current to its average value $I_0$. The distribution
  is strongly asymmetric, with the Gaussian tale at $I\gg I_0$. Inset:
  The electron occupation $f$ inside the wire as a function of the 
  rescaled coordinate $z$, under the condition that the average current 
  $I=I_0$, no current $I=0$, and large current $I=5I_0$ has been measured.}
  \label{sub-fig-test}
\end{figure}

On the classical level, the electrons in the diffusive wire are
described by the distribution function $f(z)$. Under 
transport conditions (and at zero temperature), this distribution $f(z)$
varies from $f_L=1$ in the left lead to
$f_R=0$ in the right lead. Taking the continuum limit
for the series of mesoscopic cavities,\cite{Ob2} we arrive at the action
(\ref{action-stat1}) in the form
\be S=(tI_0/e)\int\limits_{-1/2}^{1/2}dz
[-f'\lambda'+f(1-f)(\lambda')^2],
\label{action-stat3}
\ee
where we have rescaled the coordinate $z$, $\rho(z)$ has been replaced with
the distribution $f(z)$, and where $D=1$, and $F=2f(1-f)$ up to the overall
constant $I_0/e$.  This form of $F$, originating from the Pauli blocking 
factors, is quite general for fermionic systems. 
Applying now the
procedure described in the beginning of this section, we solve the
saddle-point equations and find the fields $f$ and $\lambda$,
\be
f(z,\chi)=\frac{1}{2} \left[ 1- \frac{\sinh (2\alpha z)}
{\sinh\alpha} \right]\; ,
\label{f} 
\ee
\be
\lambda (z,\chi) = 2 \, {\rm arctanh} \left[\tanh (\alpha/2) 
\tanh(\alpha z) \right]\; ,
\label{l} 
\ee
\be 
\alpha = {\rm arcsinh}\left[\sqrt{\exp(\chi)-1}\,\right],
\label{c2} 
\ee
where ${\cal H}=\alpha^2$, so that according to the Eq.\ (\ref{action-stat2})
we immediately obtain the result (\ref{ans}). 

The logarithm of the current distribution $\ln[P(I)]$ can be now found from
the equation (\ref{LNPI}). We obtain the following result: 
\be
\ln[P(I)]=-(tI_0/e)[2\alpha\coth\alpha\ln(\cosh\alpha)-\alpha^2], \ee where
$\alpha$ has to be expressed in terms of ${\cal I}=I/I_0$ by solving the
equation \be \alpha\coth\alpha=I/I_0.
\label{alpha}
\ee
The distribution $P(I)$ is strongly asymmetric around the average
current $I=I_0$ (see Fig.\ \ref{sub-fig-test}). 

In Ref.\ [1] we have proven the universality of the FCS of the transmitted
charge for a two-terminal multi-dimensional generalized wire with the noise 
tensor $F(\rho)\hat T$, being
an arbitrary function of the charge density $\rho$, and with the constant
diffusion tensor $D\hat T$. The universality means that the FCS depends
neither on the shape of the conductor, nor on its dimensionality.
The FCS of a mesoscopic wire given by Eq.\ (\ref{ans}) is a
particular example of  universal FCS. In the more general case, when $D$ 
is a function of $\rho$, the FCS depends on the geometry through only one
parameter, the geometrical conductance.

\section*{Acknowledgments}
This work was supported by the Swiss National
Science Foundation, and by INTAS 
(project 0014, open call 2001).


\begin{thebibliography}{99}
\bibitem{cap}
The physical conductance is related to the generalized 
conductance discussed here through a capacitance.
\end{thebibliography}
\end{document}